\documentclass{llncs}
\usepackage{makeidx}  
\usepackage{graphicx}
\begin{document}
\title{Weighted decomposition in high-performance lattice-Boltzmann simulations: are some lattice sites more equal than others?}
\titlerunning{Weighted decomposition}

\author{Derek Groen$^1$ \and David Abou Chacra$^1$ \and Rupert W. Nash$^1$ \and Jiri Jaros$^2$ \and Miguel O. Bernabeu$^1$ \and Peter V. Coveney$^1$}
\institute{1. Centre for Computational Science, University College London\\
20 Gordon Street, WC1H 0AJ, London, United Kingdom\\
2. Faculty of Information Technology, Brno University of Technology\\
Bozetechova 2, 612 66 Brno, Czech Republic\\
\email{d.groen@ucl.ac.uk, p.v.coveney@ucl.ac.uk}}

\maketitle              

\begin{abstract}

Obtaining a good load balance is a significant challenge in scaling up
lattice-Boltzmann simulations of realistic sparse problems to the exascale.
Here we analyze the effect of weighted decomposition on the performance of
the HemeLB lattice-Boltzmann simulation environment, when applied to sparse
domains. Prior to domain decomposition, we assign wall and in/outlet sites with
increased weights which reflect their increased computational cost. We combine 
our weighted decomposition with a second optimization, which is to sort the 
lattice sites according to a space filling curve. We
tested these strategies on a sparse bifurcation and very sparse aneurysm
geometry, and find that using weights reduces calculation load imbalance by up
to 85\%, although the overall communication overhead is higher than some of our
runs. 

\end{abstract}
\section{Introduction}

The lattice-Boltzmann (LB) method is widely applied to model fluid flow, and
relies on a stream-collision scheme applied between neighbouring points 
on a lattice. These local interactions allow LB implementations
to be efficiently parallelized, and indeed numerous high performance LB
codes exist today~\cite{Godenschwager:2013,Hasert:2013}.

Today's parallel LB implementations are able to efficiently resolve large 
non-sparse bulk flow systems (e.g., cuboids of fluid cells) using Petaflop
supercomputers~\cite{Godenschwager:2013,Groen:2011-hypo4d}. Efficiently modelling sparse
systems on large core counts is still an unsolved problem, primarily because
it is difficult to obtain a good load balance in calculation volume, neighbour
count \textit{and} communication volume for sparse geometries on large
core counts~\cite{Groen:2012}. Additionally, the presence of wall sites, 
inlets and outlets create a heterogeneity in the computational cost of 
different lattice sites. Here we test two techniques for their potential to
improve the load balance in simulations using sparse geometries, and their 
performance in general.

We perform this analysis building forth on existing advances. Indeed,
several LB codes already provide special decomposition techniques to 
more efficiently model flow in sparse geometries. For example,
Palabos~\cite{Palaboswiki}, MUSUBI~\cite{Hasert:2013} and
WaLBerla~\cite{Godenschwager:2013} apply a block-wise decomposition strategy,
while codes such as HemeLB~\cite{Groen:2012} and MUPHY~\cite{Peters:2010} rely
on third-party partitioning libraries such as ParMETIS and PT\_Scotch.

Here we implement and test a weighted decomposition technique to try and improve the
parallel simulation performance of the HemeLB simulation environment for sparse
geometries~\cite{Mazzeo:2008}, by adding weights corresponding to the
computational cost of lattice sites which do not represent bulk fluid sites.
In addition, we examine the effect of also pre-ordering the
lattice via a space-filling curve when applying this method.

Several other groups have investigated the use of weighted decomposition in
other areas, for example in environmental fluid mechanics~\cite{Barad:2009}.
In addition, Catalyurek et al.~\cite{Catalyurek:2007} investigate adaptive
repartitioning with Zoltan using weighted graphs. Specifically, Axner et
al.~\cite{Axner:2008} applied a weighting technique to a lattice-Boltzmann
solver for sparse geometries. Whereas we apply weights to vertices, they
applied heavier weights to edges near in- and outlets, to ensure that these
regions would not be distributed across several processes. 

\section{HemeLB}

\begin{figure}[!t] \centering \includegraphics[width=\textwidth]{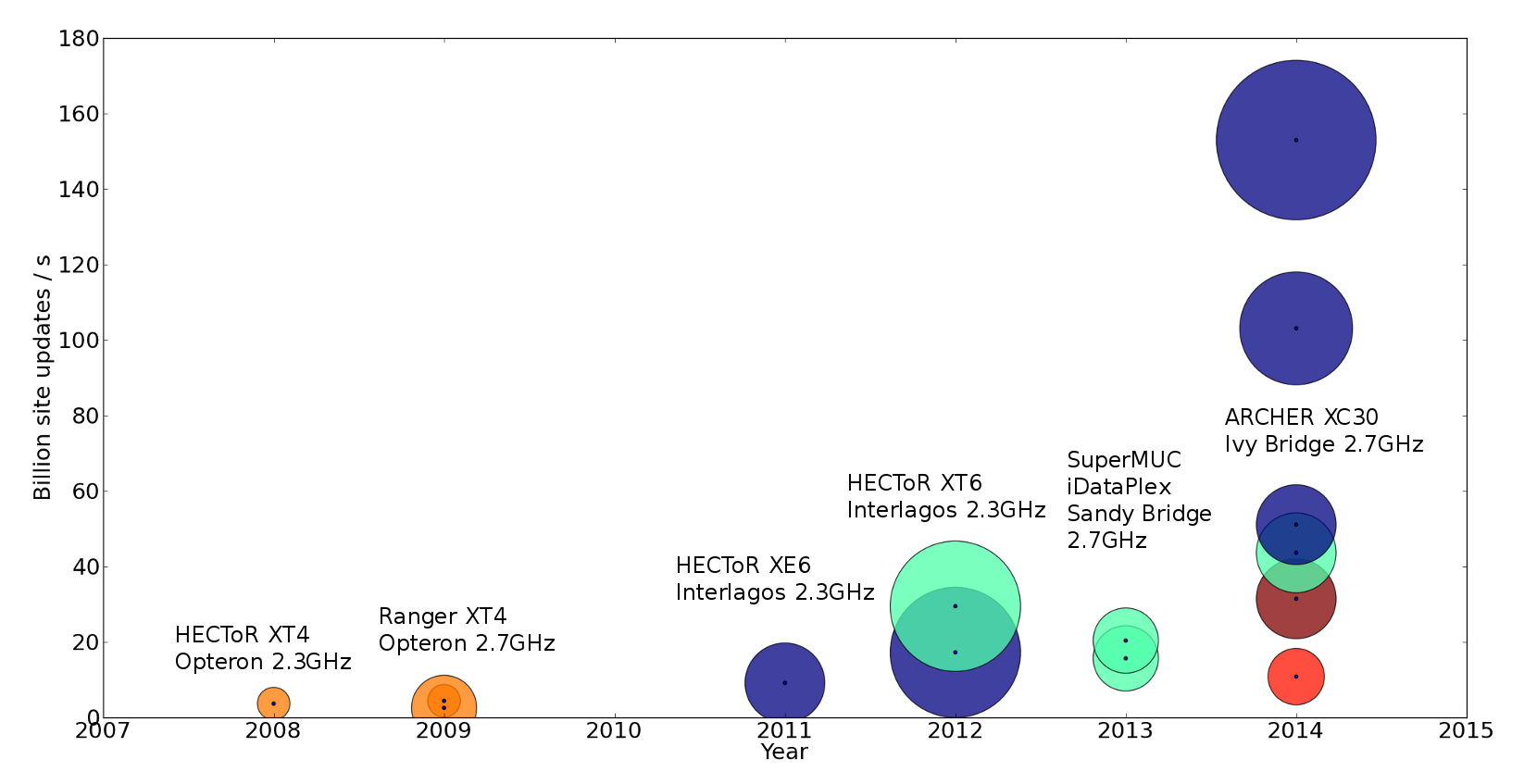}

\caption{Overview of the obtained calculation performance (in billions of lattice site 
updates per second as a function of the years in which the simulation runs were 
performed. The runs were performed on a variety of supercomputers, each of which is 
briefly described above or below the respective data points. The number of cores used
is shown by the size of the circle, ranging from 2,048 cores (smallest circles)
to 49,152 cores (largest circles). The fluid fraction is shown
by the color of the circle. These include very sparse simulation domains such as 
vascular networks (red circles), sparse domains such as bifurcations (green circles), 
ranging to non-sparse domains such cylinders (blue circles). }

\label{Fig:hemelb-hist} \end{figure}

HemeLB is a high performance parallel lattice-Boltzmann code for large scale
fluid flow in complex geometries. It is mainly written in C++.  HemeLB supports a
range of boundary conditions and collision operators~\cite{Nash:2014} and
features a streaming visualization and steering
client~\cite{Mazzeo:2010,Groen:2012}. In addition, we have equipped HemeLB with
a coupling interface, allowing it to be used as part of a multiscale
simulation~\cite{Groen:2012-2}. HemeLB uses the coalesced communication design
pattern to manage its communications~\cite{Carver:2012-2}, and relies on non-blocking
point-to-point MPI send and receive calls to perform data movements during the
simulation. We present the improvement in performance of
HemeLB over time in Fig.~\ref{Fig:hemelb-hist}. We obtained the performance data 
for this figure from a variety of sources (e.g., \cite{Mazzeo:2008,Mazzeo:2010,Groen:2012}).
Overall, the peak performance of HemeLB has improved by more than a factor 25 
between 2007 and 2014, although we do now distinguish some difference in peak 
performance between simulations with sparse geometries (e.g., aneurysm models)
and those with non-sparse geometries (e.g., cylinders). Most recently, we obtained
a performance of 153 MSUPS using 49,152 cores on the ARCHER supercomputer~\cite{Allinea}.
The geometry used in these runs was a cylinder containing 230 million lattice sites.


HemeLB originally performs decomposition in two stages, making use of the
ParMETIS graph partitioning library~\cite{parmetis-site} version 4.0.2. In the first stage it
loads the lattice arranged as blocks of 8 by 8 by 8 lattice sites. These blocks
are distributed across the processes, favoring adjacent blocks when a process
receives multiple blocks~\cite{Mazzeo:2008}. After this initial decomposition,
HemeLB then uses the {\tt ParMETIS\_V3\_PartKWay()} function to optimize the
decomposition, abandoning the original block-level structure~\cite{Groen:2012}.
This function relies on a K-way partitioning technique, which first shrinks the
geometry to a minimally decomposable size, then performs the decomposition, and
then refines the geometry back to its original size. One of the ways we can assess
the quality of the decomposition is by examining the {\em edge cut}, which is equal 
to the number of lattice neighbour links that cross process boundaries.

\section{Description of the optimizations}

We have implemented and tested two optimizations in the decomposition.

\subsection{Weighting}

\begin{figure}[!t] \centering \includegraphics[height=1.4in]{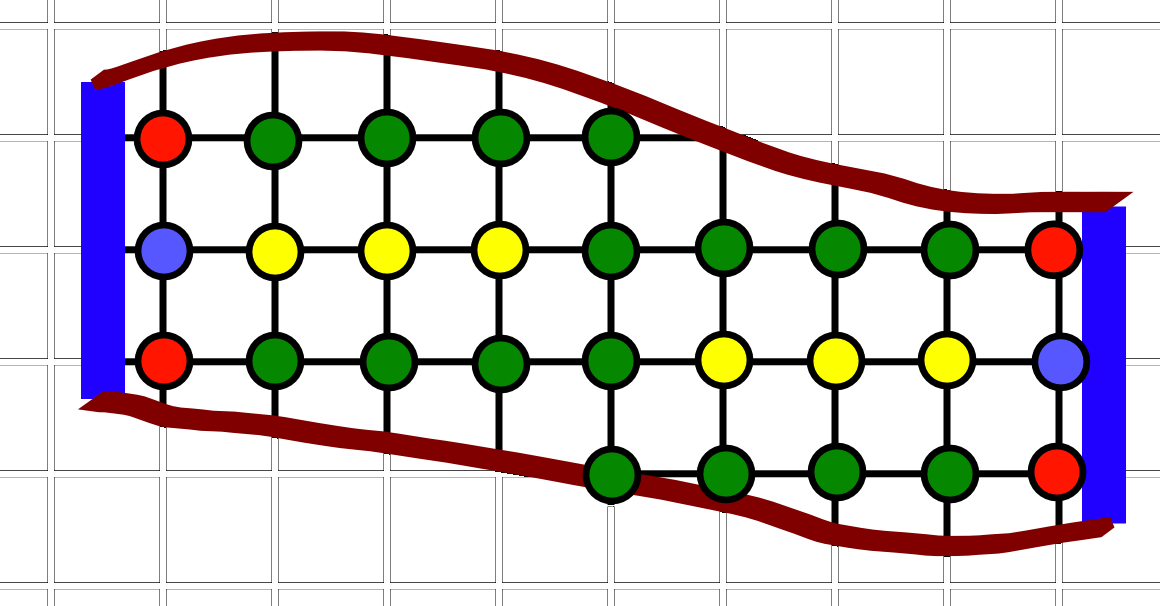}
\caption{2D example of a sparse domain with the different types of lattice
sites. In/outlets are given by the blue bars and vessel walls by the red
curves. Bulk sites are shown by yellow dots, wall sites by green dots, wall
in/outlet sites by red dots, and in/outlet sites by blue dots.}
\label{Fig:site-types} \end{figure}

Within sparse geometries, lattice-Boltzmann codes generally adopt a range of
lattice site types to encapsulate all the functionalities required to treat
flow in bulk, near walls and near in- and outlets. We provide a simple example
of a geometry containing these lattice site types in Fig.~\ref{Fig:site-types}.
By default, all types of lattice sites are weighted equally in HemeLB, which
means that graph partitioners such as ParMETIS treat all site types with equal
importance when creating a domain decomposition. However, we find that both
sites adjacent to walls and sites adjacent to in- and outlets require more
computational time to be updated. To optimize the load balance of the code, we
therefore assign heavier weights to sites which reside adjacent to wall or
in/outlet boundaries.

We are currently developing an automated tuning implementation to obtain these
computational costs at run-time. However, as a first proof of concept, we have
deduced approximate weighting values by running six simulations of cylinders
with different aspect ratios. The shorter and wider cylinders have a relatively
high ratio of in- and outlet sites, while the longer and more narrow cylinders
have a relatively high ratio of wall sites. In addition, the cylinders with an
aspect ratio near 1:1 have a relatively high ratio of bulk flow sites.

Based on these runs we have obtained estimated values for the computational
cost for each type of lattice site, by using a least-square fitting function.
We present the values of these fits, as well as rounded values we use in
ParMETIS, in Table~\ref{Tab:fits}. ParMETIS supports using weights in graphs,
provided that these weights are given as integers. As we found that using large
numbers for these weights has a negative effect on the stability of ParMETIS,
we chose to normalize and round the weightings such that bulk sites are given a
weight of 4, and the other site types are given by values relative to that base
value. Because the test runs contained only a very small number of wall +
in/outlet sites, we choose to adopt the weighting for in/outlet sites also for
the in/outlet sites which are adjacent to a wall boundary.

\begin{table}[!t]

\centering \begin{tabular}{|l|l|l|l|} \hline Site type
&\multicolumn{2}{|l|}{Obtained weight} & Rounded weight\\ \hline & Intel & AMD
& \\ \hline Bulk               &  10.0   & 10.0  & 4 \\
Wall (BFL)         &  18.708 & 20.226 & 8 \\ In/outlet          &  40.037 &
37.398 & 16 \\ Wall and in/outlet &  22.700* & 34.577* & 16 \\ \hline
\end{tabular}

\caption{Weight values as obtained from fitting against the runtimes of six
test simulations on two compute architectures (Intel SandyBridge and AMD Interlagos). 
The site type is given, followed by the weigh obtained from
fitting the performance data of the six runs, followed by the simplified
integer value we adopted in ParMETIS. In this work we use
Bouzidi-Firdaouss-Lallemand (BFL)~\cite{Bouzidi:2001} wall conditions and in
and outlet conditions described in Nash et al.~\cite{Nash:2014}. We observed
rather erratic fits for the weightings of in/outlet sites that are adjacent to
walls, as these made up only a very marginal fraction of the overall site
counts in our benchmark runs (less than 1\% in most cases).}\label{Tab:fits}

\end{table}

\subsection{Using a space-filling curve}

A second, and more straightforward, optimization we have applied is by taking
the Cartesian x,y and z coordinates of all lattice sites, and then sorting them
according to Morton-ordered space-filling curve. We do this prior to
partitioning the simulation domain, and in doing so, we effectively eliminate
any bias introduced by the early stage decomposition scheme described
in~\cite{Mazzeo:2008}. We do this by replacing the {\tt
ParMETIS\_V3\_PartKWay()} in the code function with a {\tt
ParMETIS\_V3\_PartGeomKWay()} function. This optimization is functionally independent from the weighted decomposition 
technique, but can lead to a better decomposition result from ParMETIS when 
applied.

\subsection{Other optimizations we have considered}

After having inserted these optimizations, we have also tried improving the
partition by reducing the tolerance in ParMETIS.  The
amount of load imbalance permitted within ParMETIS is indicated by the
tolerance value, and a lower value will increase the number of iterations
ParMETIS will do to reach its final state. Decreasing the
tolerance from 1.001 to 1.00001 resulted for us in an increase of the ParMETIS processing time while
showing a negligible difference in the quality of partitioning. As a result, we
have chosen not to investigate this optimization in this work.

\section{Setup}

\begin{figure}[!t] \centering \includegraphics[height=2.1in]{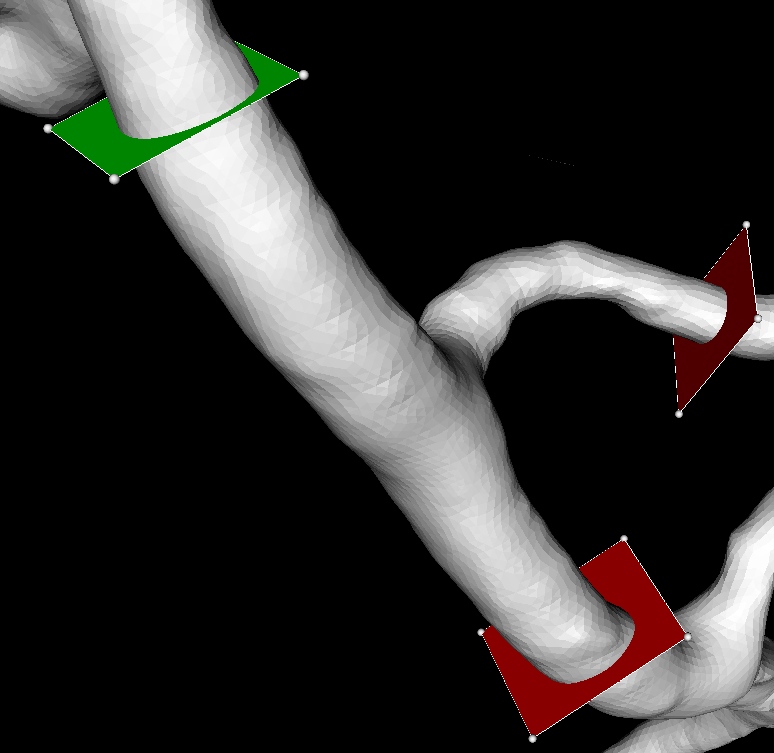}
\includegraphics[height=2.1in]{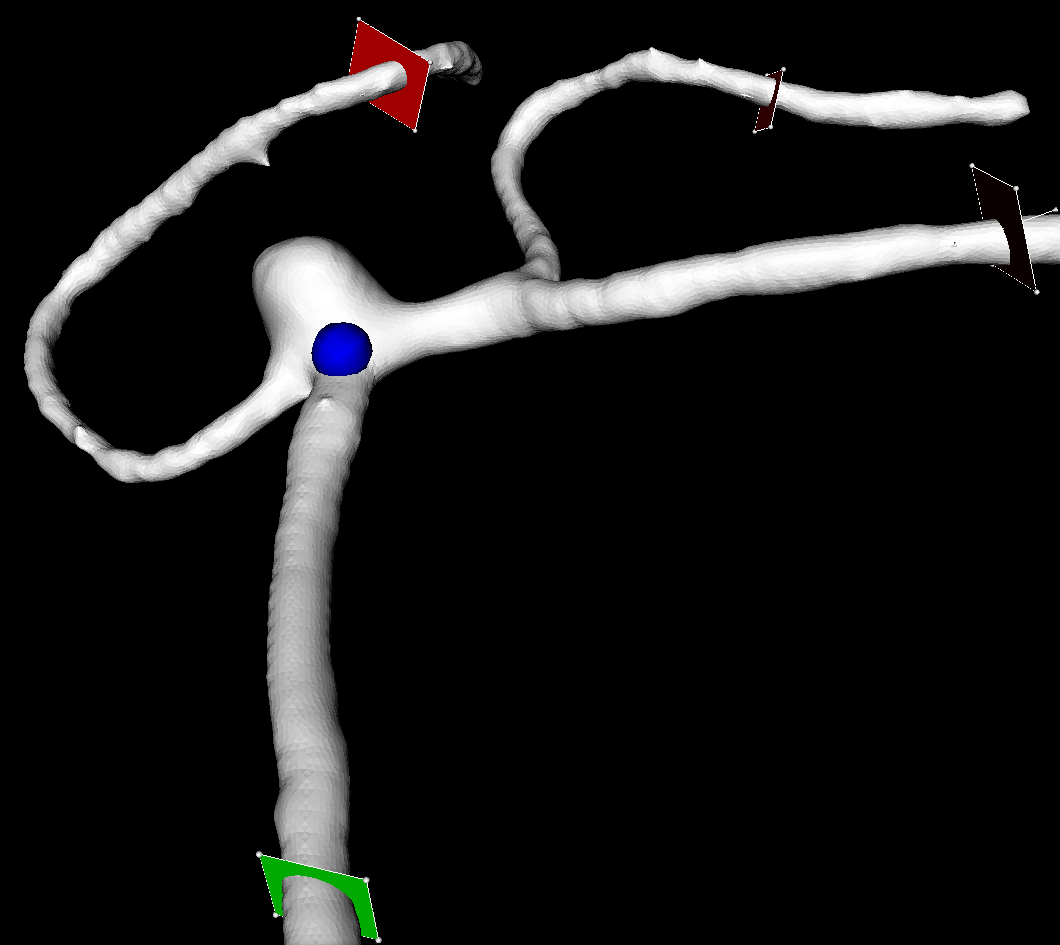} \caption{Overview of the bifurcation
geometry (left) and the aneurysm geometry (right) used in our performance
tests.  The blue blob in the aneurysm geometry is a marker indicating a region
of specific interest to the user.  The bifurcation geometry has a sparsity of
about 10\% (i.e., the lattice sites occupy about 10\% of the bounding box of
the geometry), and the aneurysm geometry a sparsity of about 1.5\%.}
\label{Fig:bifurcation} \end{figure}

In our performance tests we used two different simulation domains. These
include a smaller {\em bifurcation} geometry and a larger {\em aneurysm}
geometry (see Fig.~\ref{Fig:bifurcation} for both).  The bifurcation simulation
domain consists of 650492 lattice sites, which occupy about 10\% of the
bounding box of the geometry. The aneurysm simulation domain consists of
5667778 lattice sites, which occupy about 1.5\% of the bounding box of the
geometry. We run our simulations using pressure in- and outlets described in
Nash et al.~\cite{Nash:2014}, the LBGK collision
operator~\cite{Bhatnagar:1954}, the D3Q19 advection model and
Bouzidi-Firdaouss-Lallemand wall conditions~\cite{Bouzidi:2001}.

For our benchmarks we use the HECToR Cray XT6 supercomputer at EPCC in
Edinburgh, and compile our code using the GCC compiler version 4.3.4. We have run our 
simulations for 50000 time steps using 128 to 1024
cores for the bifurcation simulation domain, and 512 to 12288 cores for the
aneurysm simulation domain. We repeated the run for each core count five times
and averaged the results. We do this because the scheduler at HECToR does not
necessarily allocate processes within a single job to adjacent nodes; and as a
result the performance differs between runs. We have also performed several runs 
using the aneurysm simulation domain on the ARCHER Cray XC30 supercomputer at EPCC.
These runs were performed with an otherwise identical configuration. 
ARCHER relies on an Intel Ivy Bridge architecture and has a peak performance of 
about 1.6 PFLOPs in total. 

\section{Results}

\begin{figure}[!t] \centering
\includegraphics[width=2.5in]{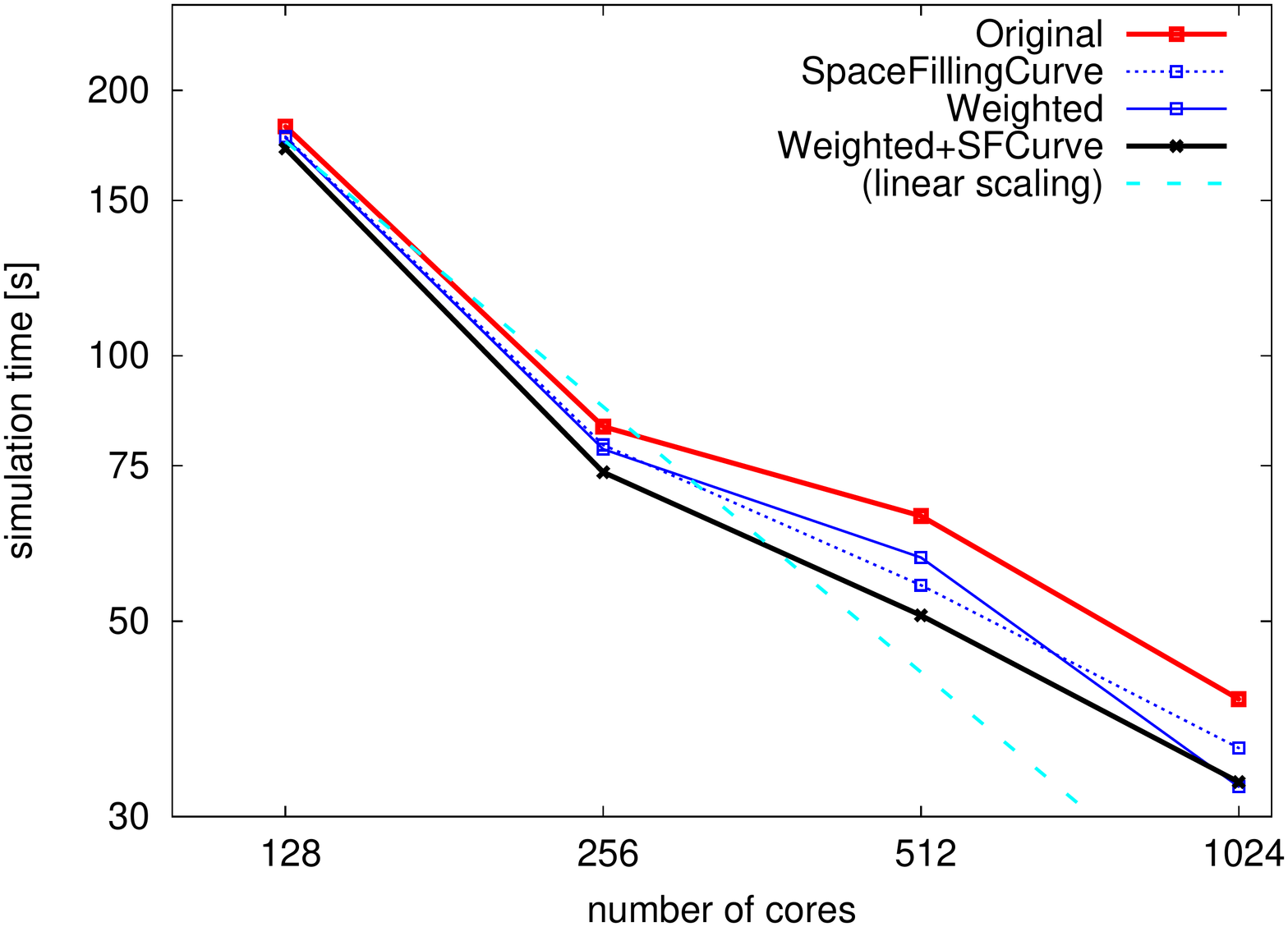}\includegraphics[width=2.5in]{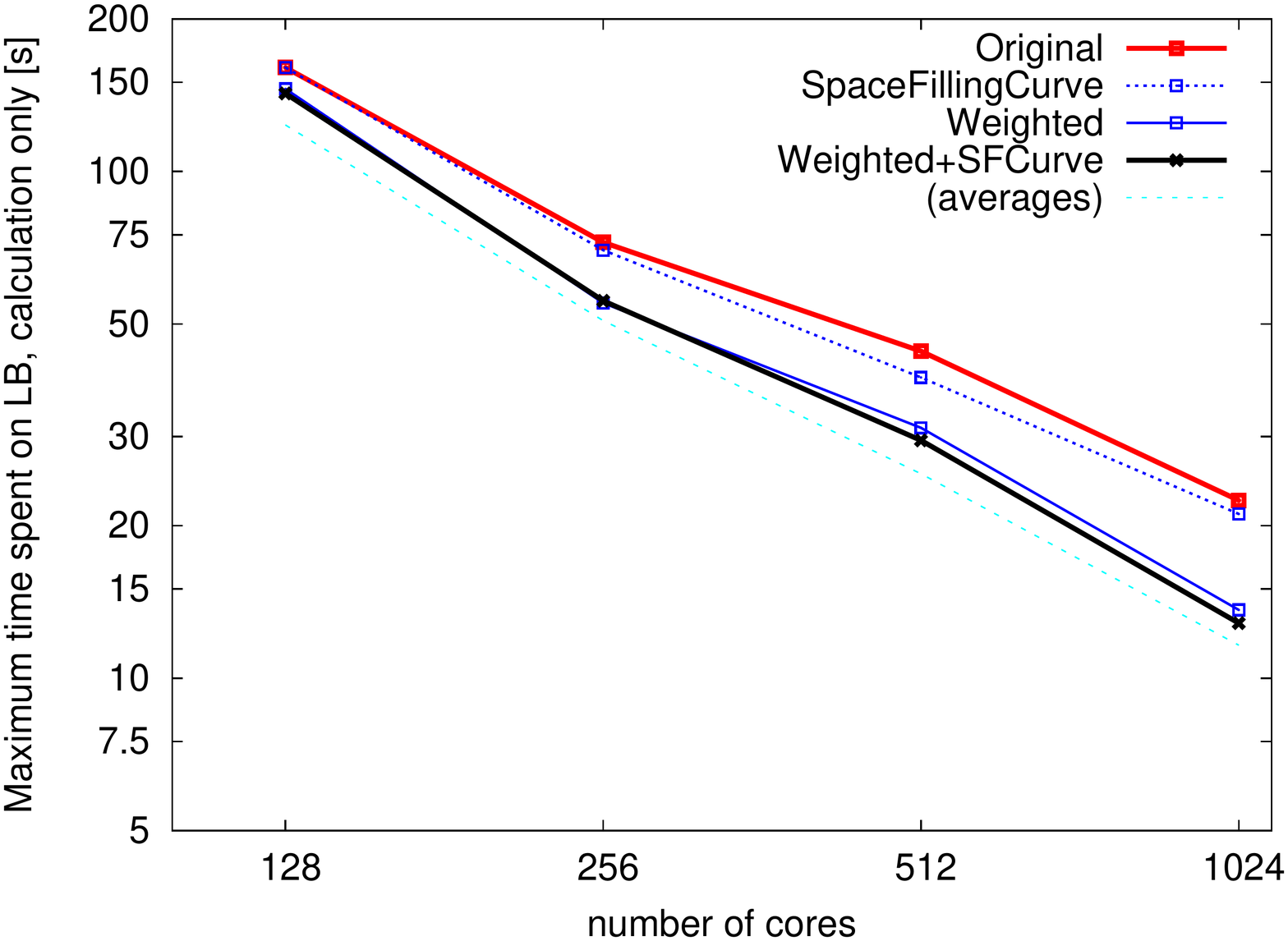}
\caption{Total simulation time and maximum LB calculation time for the simulation
using the bifurcation model, run on HECToR. We performed measurements for the non-optimized code, a
code with only weighting enabled, a code with only the space-filling curve
enabled, and a code with both enabled. We provide lines to guide the eyes. In
the image on the left we plotted a linear scaling line using a thick gray
dotted line.  In the image on the right we plotted the average LB calculation
time of all our run types using thin gray dotted lines.} \label{Fig:bsim}
\end{figure}

We present our measurements of the total simulation time and the maximum LB
calculation time for the bifurcation simulation domain in
Figure~\ref{Fig:bsim}.

We find that both incorporating a space-filling curve and using weighted
decomposition results in a reduction of the simulation time. However, the use
of a space-filling curve does little to reduce the calculation load imbalance,
whereas enabling weighted decomposition results in a reduction of the
calculation load imbalance by up to 85\%. We also examined the edge-cut
returned by ParMETIS during the domain decomposition stage. For each core
count, the edge cut obtained in all the runs was within a margin of 4.5\%, with
slightly higher edge cuts for runs using a space-filling curve or weighted
decomposition. 

\begin{figure}[!t] \centering
\includegraphics[width=2.5in]{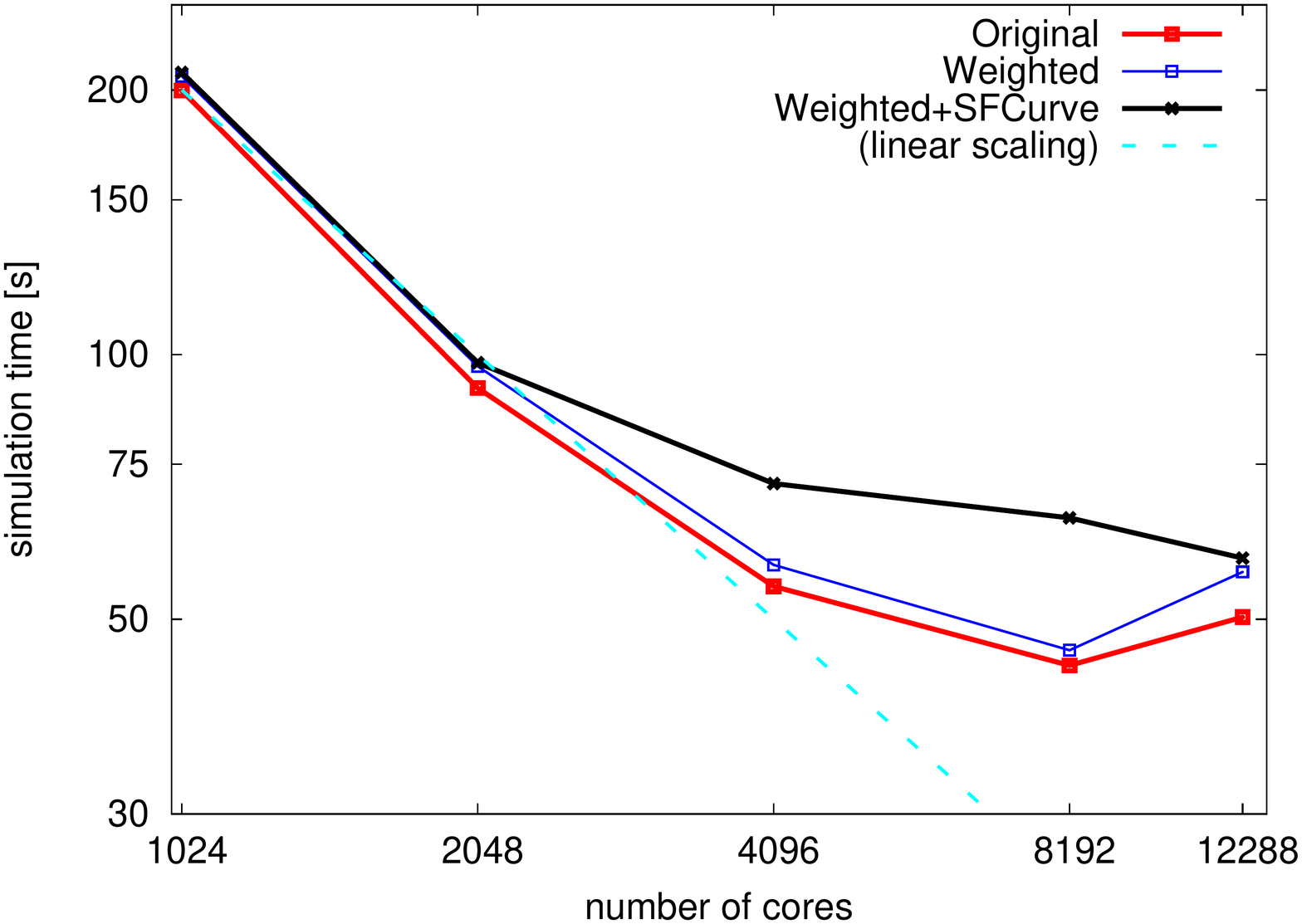}\includegraphics[width=2.5in]{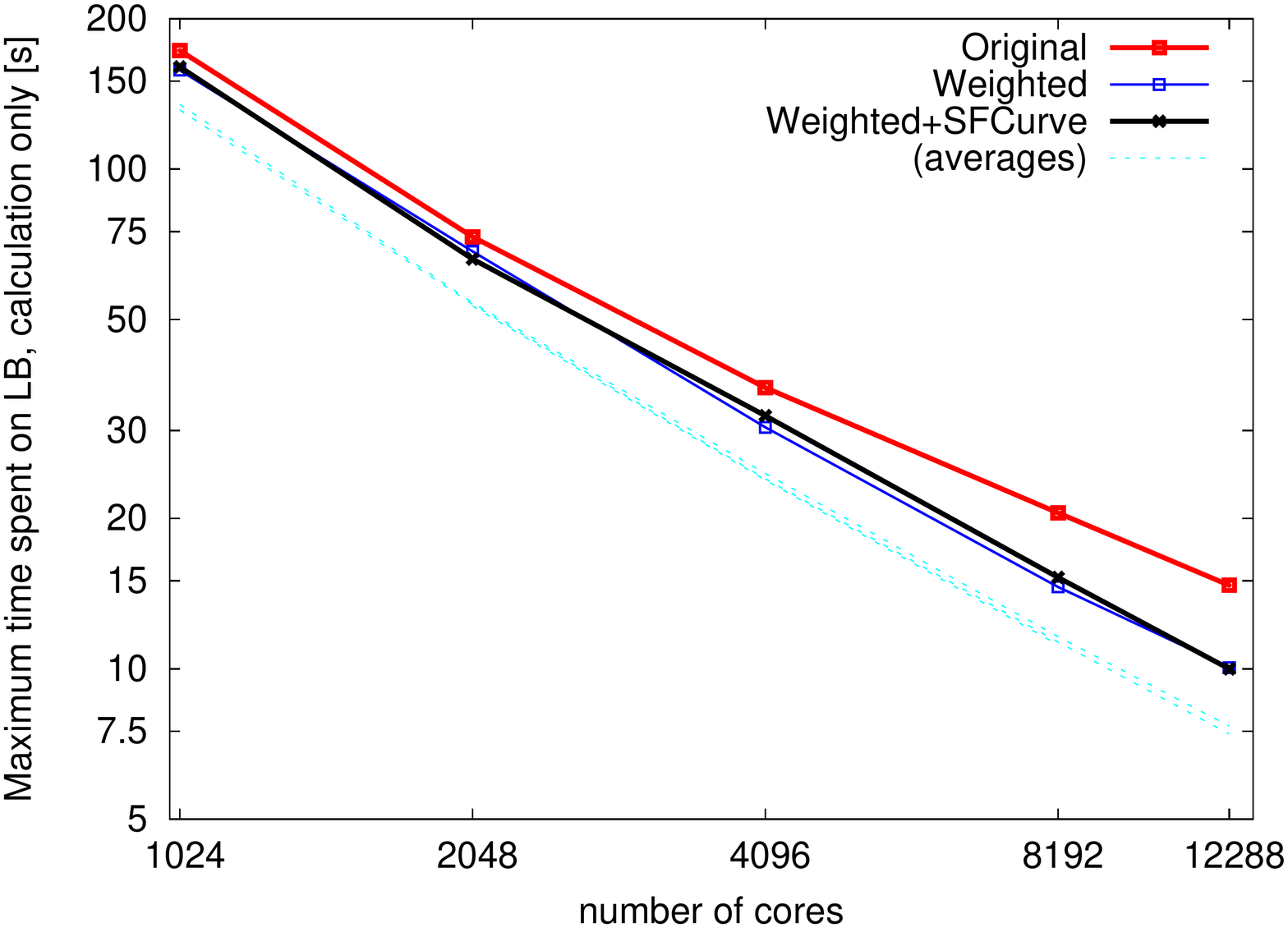}
\caption{Total simulation time and maximum LB calculation time for the simulation
of the aneurysm model, run on HECToR. See Fig:~\ref{Fig:bsim} for an explanation of the lines and symbols.
Here we only performed measurements for the non-optimized code, a code with
only the space-filling curve optimization enabled, and a code with both
optimizations enabled.} \label{Fig:asim} \end{figure}

We present our measurements of the total simulation time and the maximum LB
calculation time for the aneurysm simulation domain in Figure~\ref{Fig:asim}.
Here we find that applying weighted decomposition results in an increase of
runtime by $\sim$5\% in most of our runs. Using the space-filling curve in
addition to the weighted decomposition results in a further increase in
runtime, especially for runs performed on 4096 and 8192 cores. However, the use
of weighted decomposition also results in a calculation load imbalance which is
up to 65\% lower than that of the original simulation, while we again observe
little difference here between runs that use a space-filling curve and the runs
without. When we examine the edge cut obtained by ParMETIS in different runs,
we find that using weighted decomposition results in a slightly lower edge cut
($\sim$ 0.5\%) and using a space-filling curve results in an edge cut which is
up to 5.3\% higher.

\begin{figure}[!t] \centering
\includegraphics[width=2.5in]{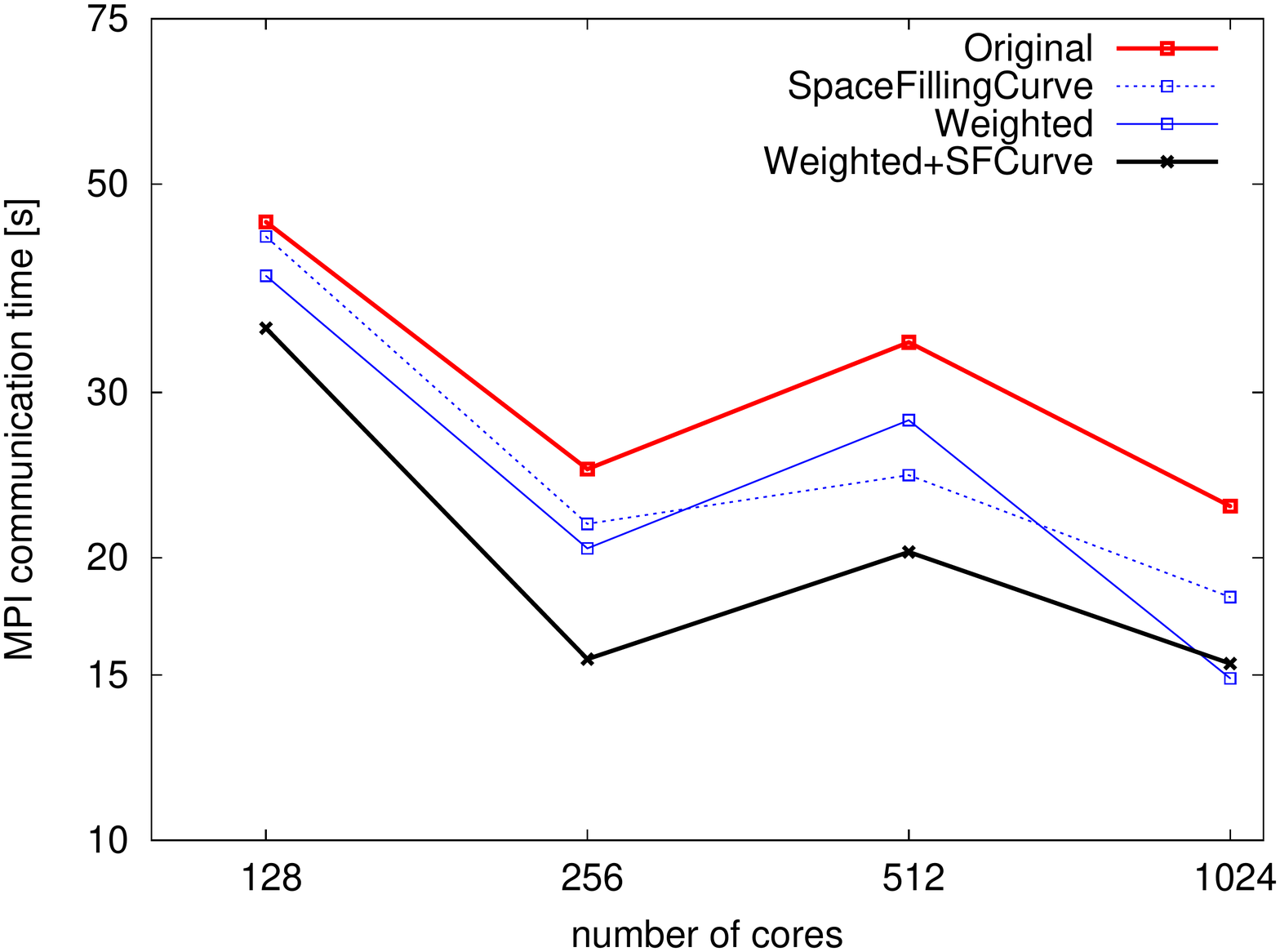}\includegraphics[width=2.5in]{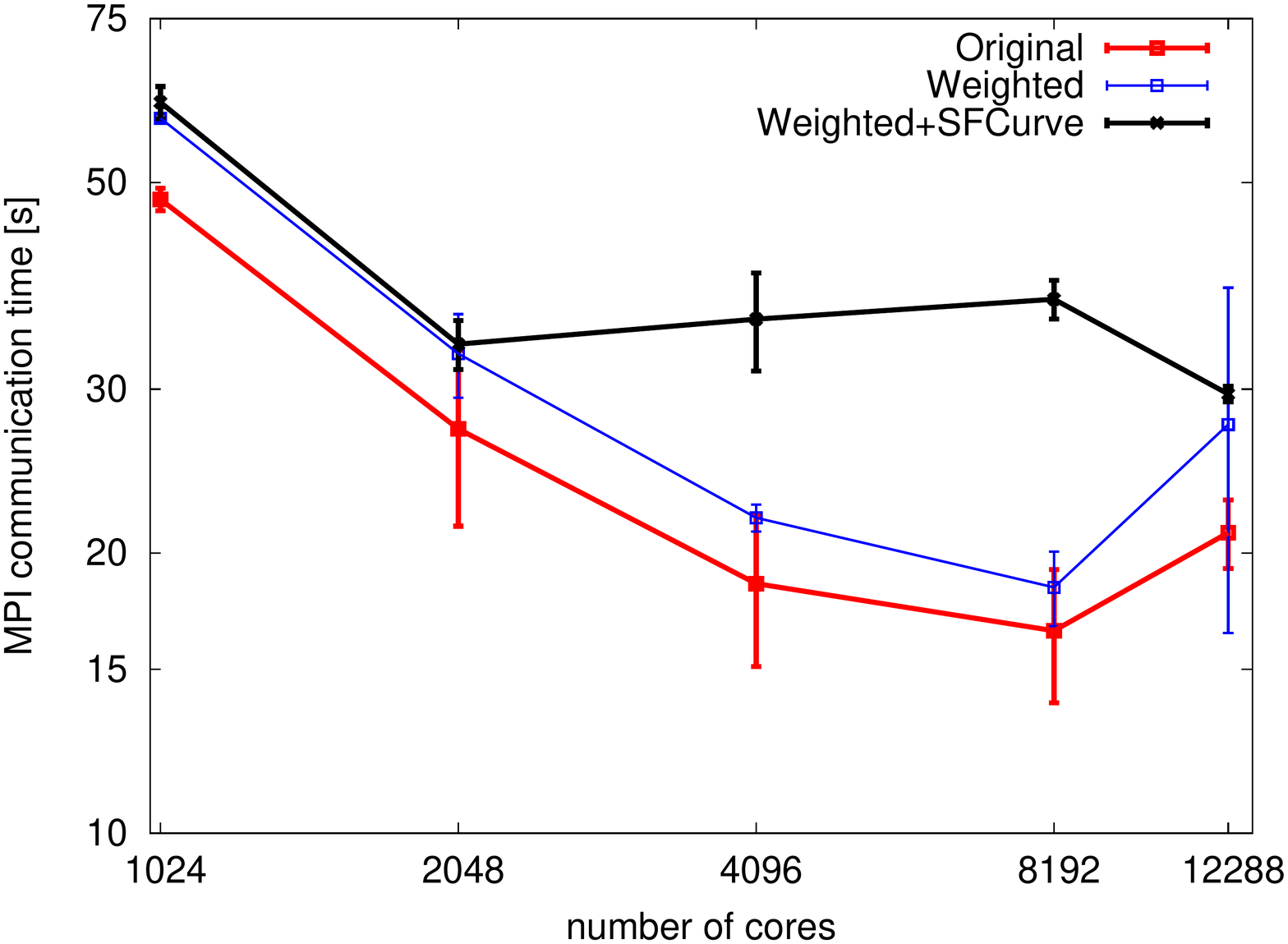}
\caption{Total MPI communication time for the simulation of the bifurcation model (left, from the run presented
in Fig.~\ref{Fig:bsim}) and the aneurysm model (right, from the run presented in Fig.~\ref{Fig:asim}).} 
\label{Fig:MPIcomm} \end{figure}

To provide more insight into the cause of the increase in simulation time, we
present our measurements of the MPI communication overhead in these runs in
Fig.~\ref{Fig:MPIcomm}. Here the runs which use our optimization strategies
take less time to do MPI communication when applied to the bifurcation
simulation domain, and more time to do MPI communications when applied to the
aneurysm domain.  These differences match largely with the differences we
observed in the overall simulation time. Because the total time spent on MPI
communications is generally larger than the calculation time for high core
counts, and the differences between the runs are considerable, the
communication performance is a major component of the overall simulation
performance. However, the communication performance correlates only weakly with
the edge cut values returned by ParMETIS and therefore the total communication
volume. For example, the slightly lower edge cut for the aneurysm simulations
with weighted decomposition is in contrast with the slightly higher
communication overhead. This means that the \textit{communication load
imbalance} is likely to be a major bottleneck in the performance of our larger
runs, and should be investigated more closely.

\subsection{Performance Results on ARCHER}

\begin{figure}[!t] \centering
\includegraphics[width=2.5in]{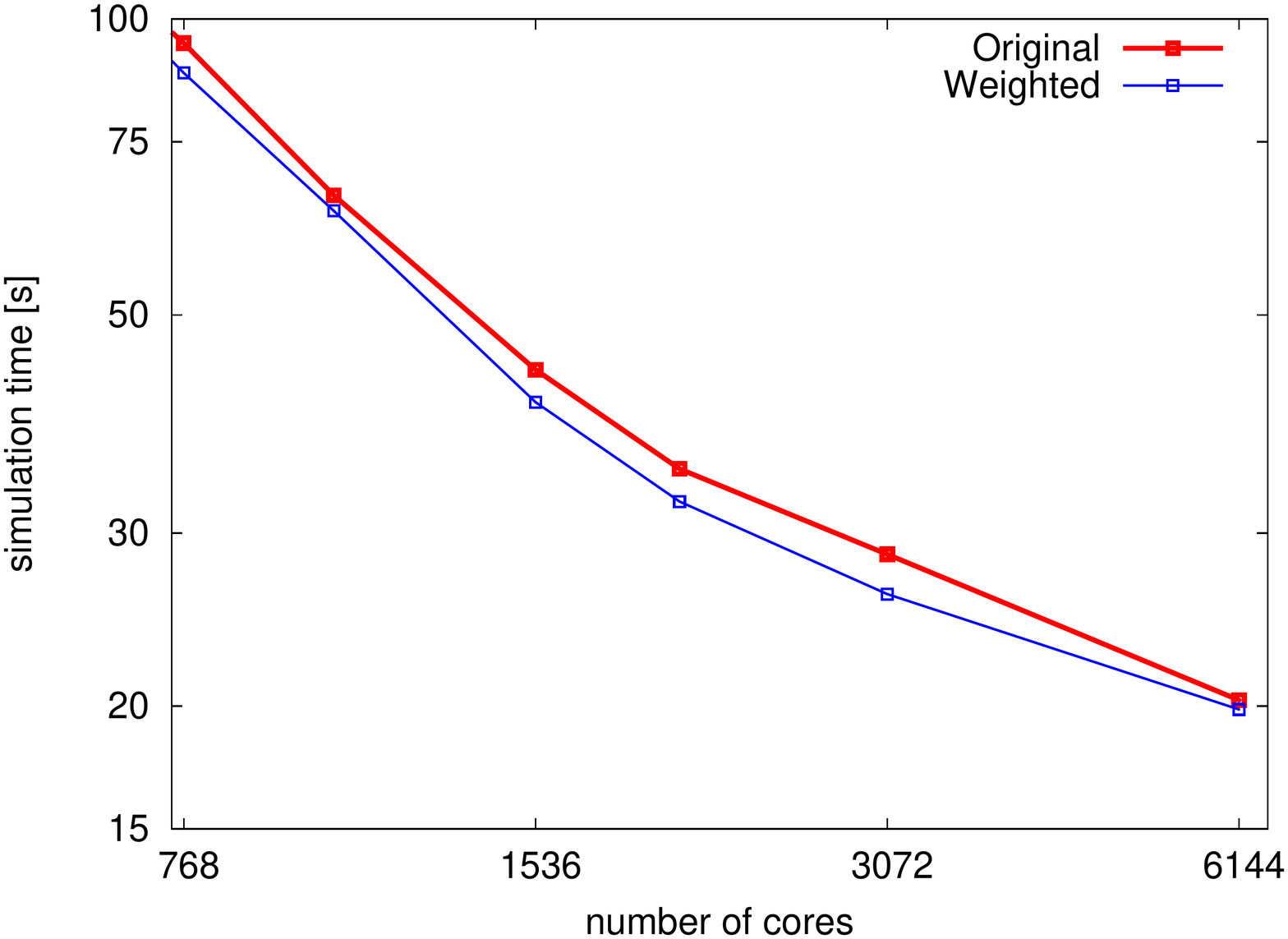}\includegraphics[width=2.5in]{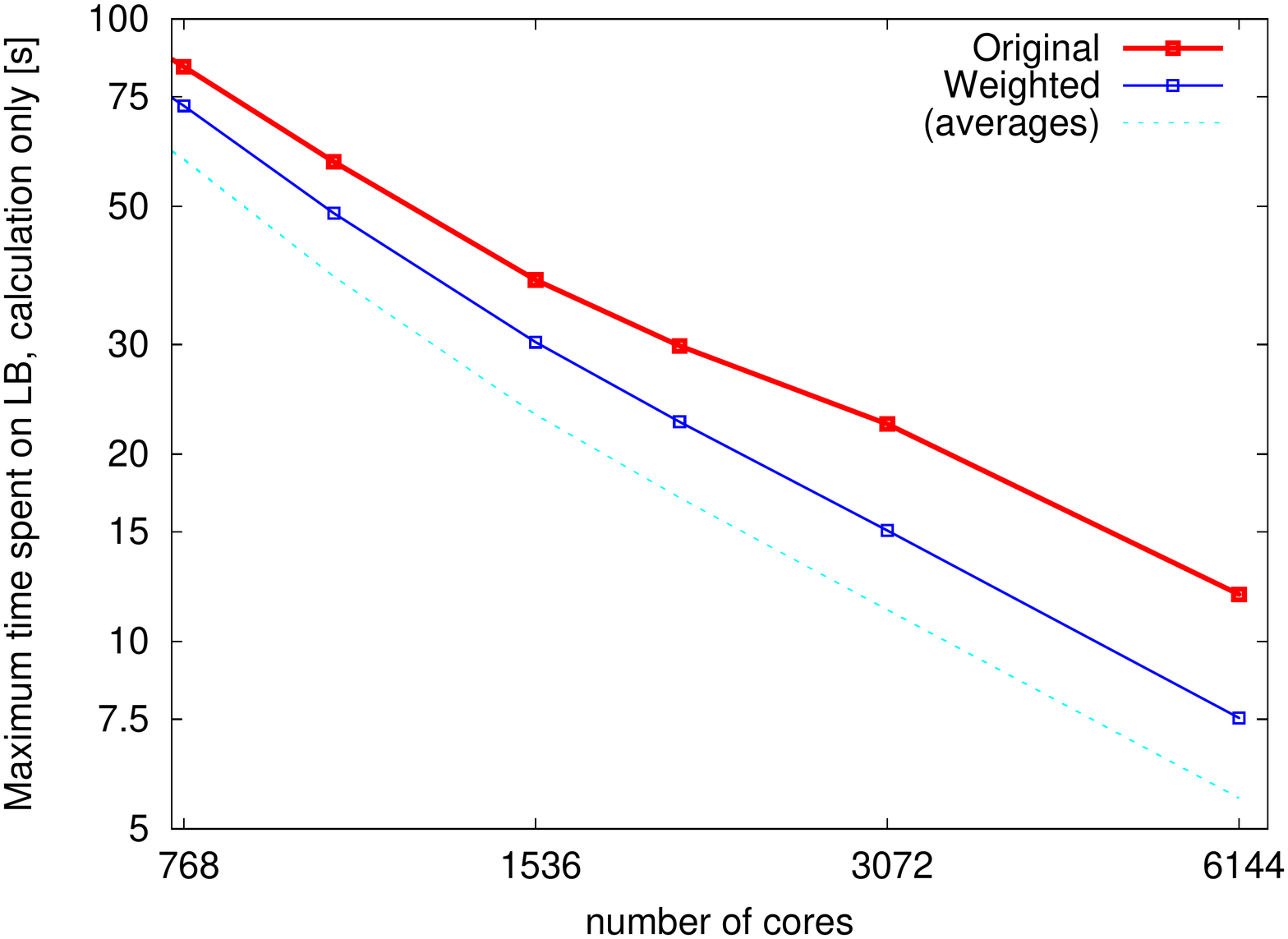}
\caption{Total simulation time and maximum LB calculation time for the simulation of the aneurysm
model, as run on ARCHER. See Fig:~\ref{Fig:bsim} for an explanation of the lines and symbols.
Here we only performed measurements for the non-optimized code, and a code with
weighted decomposition enabled.} \label{Fig:archer-asim} \end{figure}

\begin{figure}[!t] \centering
\includegraphics[width=2.5in]{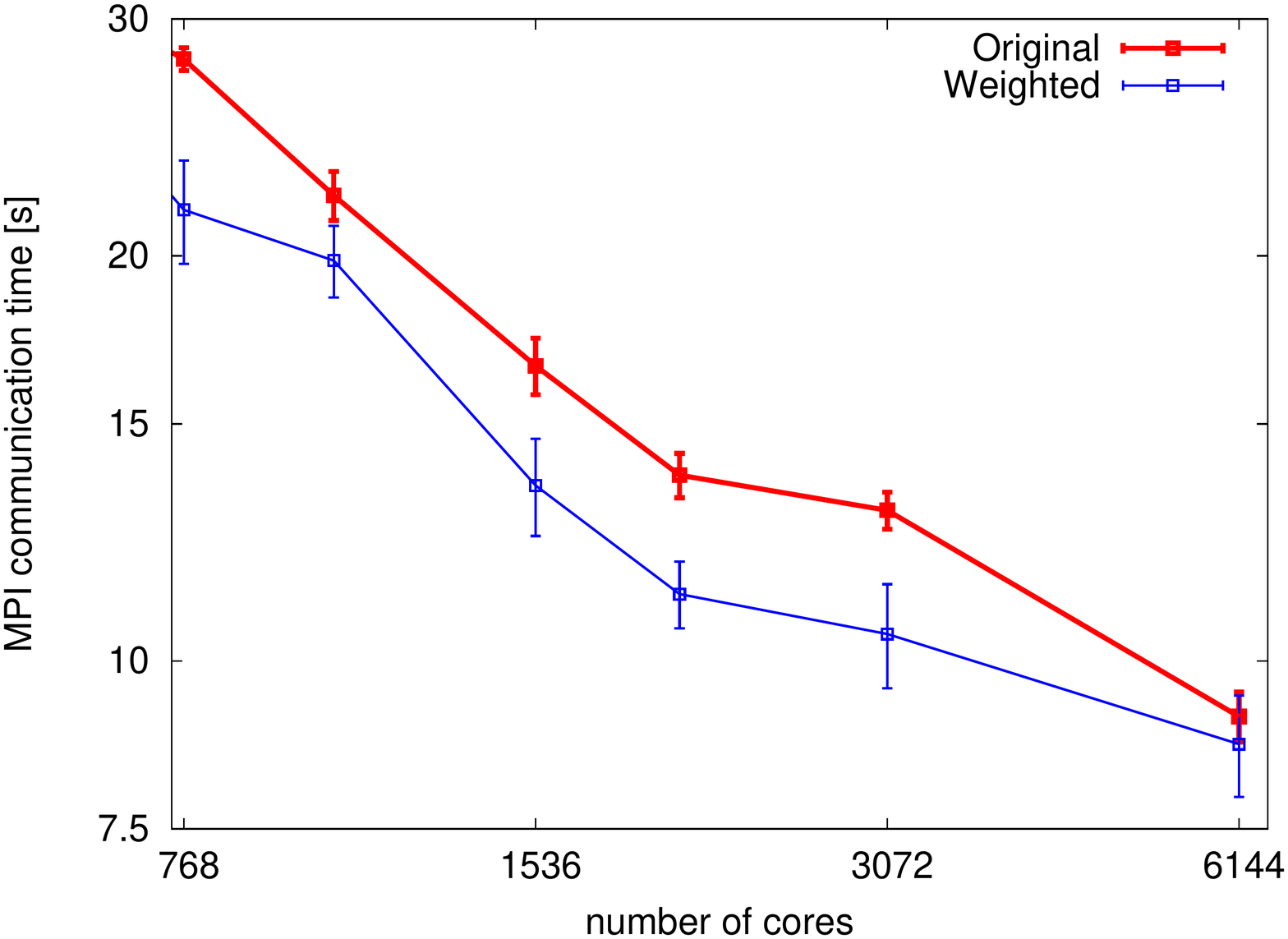}
\caption{Total MPI communication time for the run presented in Fig.~\ref{Fig:archer-asim}.} 
\label{Fig:archer-MPIcomm} \end{figure}

We have repeated the simulations using the aneurysm simulation domain on the 
ARCHER supercomputer, both with and without using weighted decomposition. 
We present the measured simulation and calculation times of these runs in Fig.~\ref{Fig:archer-asim},
and the MPI communication time in Fig.~\ref{Fig:archer-MPIcomm}.
In these runs, we obtained approximately three times the performance per core compared
to HECToR. When using weighted decomposition, the calculation load imbalance was reduced 
by up to 70\%, the simulation time by approximately 2-12\% and the MPI communication time
by approximately 5-20\%. In particular, the reduction in communication time
contrasts with the measured increase in communication time, which we observed in
the HECToR runs. This difference could be attributed to the superior
network architecture of ARCHER, and/or the large memory per core, which may
have resulted in ParMETIS reaching a domain decomposition with better communication
load balance. 

\section{Discussion and conclusions}

We presented an approach for weighted decomposition and assessed its effect on the
performance of the HemeLB bloodflow simulation environment. The use of lattice 
weights in our decomposition scheme provides
the strongest improvement in calculation load balance, and delivers an
improvement in the simulation performance for the bifurcation geometry.  
However, the use of weighted decomposition (both with and without the space-filling curve
optimization) sometimes results in a higher communication overhead of the
aneurysm simulations, despite negligible changes in the communication volume.
Indeed, for these blood flow simulations it appears that a low edge cut is only
a minor factor in the overall communication performance for sparse problems,
even though graph partitioning libraries are frequently optimized to accomplish
such a minimal edge cut. This is in accordance with some earlier conclusions in
the literature~\cite{Hendrickson:2000}. We intend to more thoroughly investigate the
communication load imbalance of our larger runs. As part of preparing HemeLB
for the exascale within the CRESTA project, we are working with experts from
the Deutschen Zentrums f\"{u}r Lucht und Raumfahrt (DLR) to enable domain
decompositions using PT-Scotch and Zoltan. The use of these alternate graph
partitioning libraries may result in further performance improvements,
especially if these libraries optimize not only for a calculation load balance
and a low edge cut, but also take into account other communication
characteristics. Furthermore, since we have observed differences in site weights between
different computer architectures, we are looking into an "auto-tuning" function
that automatically calculates the weights at runtime or compilation time.

\section{Acknowledgements}

We thank Timm Krueger for his valuable input.
This work has received funding from the CRESTA and MAPPER projects within the
EC-FP7 (ICT-2011.9.13) under Grant Agreements nos. 287703 and 261507, and 
from EPSRC Grants EP/I017909/1 (www.2020science.net) and EP/I034602/1. This 
work made use of the HECToR supercomputer at EPCC in Edinburgh, funded by the 
Office of Science and Technology through EPSRC's High End Computing Programme.

\bibliographystyle{plain}
\bibliography{Library}

\end{document}